\documentclass[a4paper,oneside,english]{amsart}
\usepackage[T1]{fontenc}
\usepackage[latin9]{inputenc}
\usepackage{listings}
\usepackage{babel}
\usepackage{float}
\usepackage{amsthm}
\usepackage{amstext}
\usepackage{amssymb}
\usepackage{graphicx}
\usepackage[unicode=true,
 bookmarks=true,bookmarksnumbered=false,bookmarksopen=false,
 breaklinks=false,pdfborder={0 0 1},backref=false,colorlinks=false]
 {hyperref}
\hypersetup{pdftitle={The block cipher NSABC (public domain)},
 pdfauthor={Alice Nguyenova-Stepanikova & Tran Ngoc Duong},
 pdfsubject={Cryptography},
 pdfkeywords={block ciphers, algebaic, tweakable, IDEA, Skipjack}}
\usepackage{breakurl}

\makeatletter

\floatstyle{ruled}
\newfloat{algorithm}{tbp}{loa}
\providecommand{\algorithmname}{Algorithm}
\floatname{algorithm}{\protect\algorithmname}

\numberwithin{equation}{section}
\numberwithin{figure}{section}
\newenvironment{lyxlist}[1]
{\begin{list}{}
{\settowidth{\labelwidth}{#1}
 \setlength{\leftmargin}{\labelwidth}
 \addtolength{\leftmargin}{\labelsep}
 }}
{\end{list}}
\newenvironment{lyxcode}
{\par\begin{list}{}{
\setlength{\rightmargin}{\leftmargin}
\setlength{\listparindent}{0pt}% needed for AMS classes
\raggedright
\setlength{\itemsep}{0pt}
\setlength{\parsep}{0pt}
\normalfont\ttfamily}%
 \item[]}
{\end{list}}

\makeatother

\begin{document}

\title{THE BLOCK CIPHER NSABC (Public domain)}

\author{Alice Nguyenova-Stepanikova ({*})}

\thanks{%
\thanks{({*}) Hradcany, Praha, Czech Republic%
}}

\author{Tran Ngoc Duong ({*}{*})}

\thanks{%
\thanks{({*}{*}) Pernink, Karlovy Vary, Czech Republic. E-mail: \texttt{tranngocduong@gmail.com}%
}}

\date{May 8th, 2011. This is version 2 of the algorithm, superseding version
1 that was published on Usenet July 2010.}
\begin{abstract}
We introduce NSABC/$w$ \textendash{} Nice-Structured Algebraic Block
Cipher using $w$-bit word arithmetic, a $4w$-bit analogous of Skipjack
\cite{NSA98} with $5w$-bit key. The Skipjack's internal 4-round
Feistel structure is replaced with a $w$-bit, 2-round cascade of
a binary operation $(x,z)\mapsto(x\boxdot z)\lll(w/2)$ that permutes
a text word $x$ under control of a key word $z$. The operation $\boxdot$,
similarly to the multiplication in IDEA \cite{LM91,LMM91}, bases
on an algebraic group over $w$-bit words, so it is also capable of
decrypting by means of the inverse element of $z$ in the group. The
cipher utilizes a secret $4w$-bit tweak \textendash{} an easily changeable
parameter with unique value for each block encrypted under the same
key \cite{LRW02} \textendash{} that is derived from the block index
and an additional $4w$-bit key. A software implementation for $w=64$
takes circa 9 clock cycles per byte on x86-64 processors.
\end{abstract}


\maketitle

\section{Introduction}

In the today's world full of crypto algorithms, one may wonder what
makes a block cipher attractive.

In the authors' opinion, the answer to the question is one word: elegance.
If something looks nice, then there is a big chance that it is also
good.

An elegant specification makes it easier to memorize. Memorability
makes it easier to realize and to analyze, that allows for fruitful
cryptanalytic results, leading to deeper understanding which, in turn,
makes greater confidence in the algorithm.

The elegance comprises the following features: 
\begin{itemize}
\item Few algebraic operations. Using of many operations results in hardly-tractable
and possibly undesirable interactions between them.
\item Simple and regular key schedule. A complex key schedule, which effectively
adds another, unrelated, function to the cipher, results in hardly-tractable
and possibly undesirable interactions between the functions.
\end{itemize}
IDEA, a secure block cipher designed by Xuejia Lai and James L. Massey
\cite{LM91,LMM91} is an example of elegance. Besides being elegant
with an efficient choice and arrangement of algebraic operations,
it is elegant for some more features: 
\begin{itemize}
\item The use of incompatible group operations, where \emph{incompatible}
means there are no simple relations (such as distributivity) between
them. The incompatibility eliminates any exploitable algebraic property
thus makes it infeasible to solve the cipher algebraically.
\item The use of modular multiplication. Multiplication produces huge mathematical
complexity while consuming few clock cycles on modern processors.
It thus greatly contributes to security and efficiency of the cipher.
\end{itemize}
However, IDEA uses multiplication modulo the Fermat prime $2^{w}+1$
which does not exist for $w=32$ or $w=64$, making it not extendable
to machine word lengths nowadays. Furthermore, its key schedule is
rather irregular due to the rotation of the primary key.

Skipjack, a secure block cipher designed by the U.S. National Security
Agency \cite{NSA98}, is another example of elegant design. Besides
being elegant with an efficient, simple and regular key schedule,
it is elegant for one more feature: the use of two ciphers \textemdash{}
an outer cipher, or \emph{wrapper}, consisting of first and last rounds,
and an inner cipher, or \emph{core}, consisting of middle rounds.

The terms {}``core'' and {}``wrapper'' were introduced in the
design rationale of a structural analogous of Skipjack: the block
cipher MARS \cite{IBM98}. MARS's designers justify this two-layer
structure by writing that it breaks any repetitious property, it makes
any iterative characteristic impossible, and it disallows any propagation
of eventual vulnerabilities in either layer to the other one, thus
making attacks more difficult. The wrapper is primarily aimed at fast
diffusion and the core primarily at strong confusion. As Claude E.
Shannon termed in his pioneer work \cite{Sha49}, \emph{diffusion}
here refers to the process of letting each input bit affect many output
bits (or, equivalently, each output bit be affected by many input
bits), and \emph{confusion} here refers to the process of letting
that affection very involved, possibly by doing it multiple times
in very different ways. If a cipher is seen as a polynomial map in
the plaintext and the key to the ciphertext, then the methods of diffusion
and confusion can be described as the effort of making the polynomials
as complete as possible, i.e. such that they contain virtually all
terms at all degrees. This \emph{algebraic} approach is very evident
in the structure of Skipjack (see Figure~\ref{fig:The-full-cipher-2}).
Skipjack (as opposed to MARS) was moreover sought elegant as the wrapper
there is, in essence, the inverse function of the core.

However, Skipjack uses an S-box that renders it rather slow, hard
to program in a secure and efficient manner, and not extendable to
large machine word lengths, as such.

This article describes an attempt to combine the elegant idea of using
incompatible and complex machine-oriented algebraic operations in
IDEA with the elegant structure of Skipjack into a scalable and tweakable
block cipher called NSABC \textemdash{} Nice-Structured Algebraic
Block Cipher.

NSABC is scalable. It is defined for every even word length \emph{w}.
It encrypts a 4\emph{w}-bit text block under a 5\emph{w}-bit key,
thus allows scaling up with 8-bit increment in block length and 10-bit
increment in key length.

NSABC is tweakable. It can use an easily changeable 4\emph{w}-bit
parameter, called \emph{tweak} \cite{LRW02}, to make a unique version
of the cipher for every block encrypted under the same key. Included
in the specification is a formula for changing the tweak.

NSABC makes use of entirely the overall structure of Skipjack, including
the key schedule, and only replaces the internal 4-round Feistel structure
of Skipjack with another structure. The new structure consists of
two rounds of the binary operation $(x,z)\mapsto(x\underset{e}{\boxdot}z)\lll(w/2)$,
that encrypts a text word \emph{x} using a key word \emph{z} and a
key-dependent word \emph{e}. The operation $\underset{e}{\boxdot}$
is derived from an algebraic group over \emph{w}-bit words taking
$e$ as the unit element, so it is also capable of decrypting by means
of the inverse element of \emph{z} in the group. The two rounds are
separated by an exclusive-or (XOR) operation that modifies the current
text word by a tweak word.

NSABC is put in public domain. As it bases on Skipjack, eventual users
should be aware of patent(s) that may be possibly held by the U.S.
Government and take steps to make sure the use is free of legal issues.
We (the designers of NSABC) are not aware of any patent related to
other parts of the design.

The rest of the article is organized as follows. Section 2 defines
operations and notations. Section 3 specifies the cipher. Section
5 gives numerical examples. Section 4 suggests some implementation
techniques. Section 6 concludes the article. Source code of software
implementations are given in the Appendices.

\section{Definitions}

\subsection{Operations on words}

Throughout this article, \emph{w} denotes the machine word length.
We use the symbols $\boxplus$, $\boxminus$, $\boxtimes$ and $(.)^{-1}$
to denote addition, subtraction (and arithmetic negation), multiplication
and multiplicative inversion, respectively, modulo $2^{w}$ (unless
otherwise said). We use the symbols $\neg$ and $\oplus$ to denote
bit-wise complement and exclusive-or (XOR)\emph{ }on \emph{w}-bit
operands (unless otherwise said). We write $x\lll n$ to denote leftward
rotation (i. e. cyclic shift toward the most significant bit) of \emph{x},
that is always a \emph{w}-bit word, by \emph{n} bits. For even \emph{w},
the symbol $(.)^{\mathrm{S}}$ denotes swapping the high and low order
halves, i.e. $x{}^{\mathrm{S}}=x\lll(w/2).$

Let's define binary operation $\odot$ by

\[
x\odot y=2xy\boxplus x\boxplus y
\]

and binary operation $\boxdot$ by

\[
x\boxdot y=2xy\boxplus x\boxminus y
\]

The bivariate polynomials on the right hand side are permutation polynomials
in either variable for every fixed value of the other variable {[}Riv99{]}.
In other words, $\odot$ and $\boxdot$ are quasi-group operations.

Furthermore, $\odot$ is a group operation over the set of \emph{w}-bit
numbers%
\footnote{This fact, although simple and straightforward, does not seem to have
been mentioned in the literature.%
}. This fact becomes obvious by considering an alternative definition
for the $\odot$ operation \cite{Mey97}: it can be done by dropping
the rightmost bit, which is always {}``1'', of the product modulo
$2^{w+1}$ of the operands each appended with an {}``1'' bit. Symbolically,

\[
x\odot y=\left[(2x+1)(2y+1)-1\right]/2\;(\mathrm{\textrm{mod}\:}2^{w})
\]

The group defined by $\odot$ is thus isomorphic to the multiplicative
group of odd integers modulo $2^{w+1}$, via the isomorphism 
\[
x\mapsto2x+1
\]

The unit (i.e., identity) element of the group is 0. The inverse element
of \emph{x}, denoted $\bar{x}$, is 

\[
\bar{x}=\boxminus x(2x\boxplus1)^{-1}
\]

The following relations are obvious. 

\[
x\odot y=\boxminus\left[(\boxminus x)\boxdot y\right]
\]

\[
x\boxdot y=\boxminus\left[(\boxminus x)\odot y\right]
\]

Since the unary operator $\boxminus$ is an involution, the following
relations hold.

\[
(x\boxdot y)\boxdot z=x\boxdot(y\odot z)
\]

\[
x\boxdot0=0
\]

Notice that 0 is the right unit element w.r.t. the operation $\boxdot$.
Hence

\[
(x\boxdot y)\boxdot\bar{y}=x
\]
which means that $\bar{y}$ is also the right inverse element of \emph{y}
w. r. t. the $\boxdot$ operation. 

Since $(\neg x)\boxplus x=\boxminus1$ holds for every \emph{x}, the
following relations hold.

\begin{gather*}
(\neg x)\odot y=\neg(x\odot y)
\end{gather*}

\[
(1\boxminus x)\boxdot y=1\boxminus(x\boxdot y)
\]

Let \emph{e} be a fixed \emph{w}-bit number. Let's define binary operations
$\underset{e}{\odot}$ and $\underset{e}{\boxdot}$ by 

\[
x\underset{e}{\odot}y=(x\boxminus e)\odot(y\boxminus e)\boxplus e=2xy\boxplus(1-2e)(x+y-e)
\]

\[
x\underset{e}{\boxdot}y=(x\boxplus e)\boxdot(y\boxminus e)\boxminus e=2xy\boxplus(1-2e)(x-y+e)
\]

Then $\underset{e}{\odot}$ and $\underset{e}{\boxdot}$ are quasi-group
operations over the set of \emph{w}-bit numbers. This follows from
a more general fact that the right-hand side trivariate polynomials
are permutations in either variable while keeping the other two fixed
{[}Riv99{]}. Actually, the symbols $\underset{e}{\odot}$ and $\underset{e}{\boxdot}$
each defines an entire family of binary operations, of which each
is uniquely determined by \emph{e}.

Furthermore, from the definition it immediately follows that $\underset{e}{\odot}$
is a group operation, namely, the group is isomorphic to one defined
by $\odot$ via the isomorphism

\[
x\mapsto x\boxminus e
\]

The unit element of the group is \emph{e}. 

The inverse element of \emph{x} in the group, denoted $\frac{e}{x}$,
is

\[
\frac{e}{x}=\overline{x\boxminus e}\boxplus e=\left[(2e-1)x\boxminus2e(e-1)\right]\,\boxtimes\,\left[2(x-e)\boxplus1\right]^{-1}
\]

Simple calculation proves the following relations.

\[
x\underset{e}{\odot}y=\boxminus\left[(\boxminus x)\underset{e}{\boxdot}y\right]
\]

\[
x\underset{e}{\boxdot}y=\boxminus\left[(\boxminus x)\underset{e}{\odot}y\right]
\]

\[
(1\boxminus2e\boxminus x)\underset{e}{\boxdot}y=1\boxminus2e\boxminus(x\underset{e}{\boxdot}y)
\]

\[
(x\underset{e}{\boxdot}y)\underset{e}{\boxdot}z=x\underset{e}{\boxdot}(y\underset{e}{\odot}z)
\]

\[
(x\underset{e}{\boxdot}y)\underset{e}{\boxdot}\frac{e}{y}=x
\]

Notice that \emph{e} is also the right unit element w. r. t. $\underset{e}{\boxdot}$,
and $\frac{e}{y}$ also the right inverse element of \emph{y} w. r.
t. $\underset{e}{\boxdot}$. The operation $\underset{e}{\boxdot}$,
which is non-commutative and non-associative, will be used for encryption
and, due to the existence of right inversion, also for decryption.

\subsection{Order notations}

We write multi-part data values in \emph{string} (or \emph{number})
notation or \emph{tuple} (or \emph{vector}) notation. In string notation,
the value is written as a sequence of symbols, possibly separated
by space(s) that are insignificant. In tuple notation, the value is
written as a sequence, in parentheses, of comma-separated symbols.

For examples, \emph{z y x} and 43 210 are in string notation, $(x,y,z)$
and $(0,1,2,3,4)$ are in tuple notation.

The string notation indicates \emph{high-first} order: the first (i.e.
leftmost) symbol denotes the most significant part of the value when
it is interpreted as a number.

Conversely, the tuple notation indicates \emph{low-first} order: the
first symbol denotes the least significant part of the value when
it is interpreted as a number.

For examples, to interpret a 3-word number, $x_{2}$ denotes the most
significant word of $x_{2}x_{1}x_{0}$ and $x_{0}$ denotes the least
significant word of $(x_{0},x_{1},x_{2})$.

The same value may appear in either notation. Thus, for example, for
every \emph{a}, \emph{b}, \emph{c} and \emph{d},

\[
a\: b\: c\: d=(d,c,b,a)
\]

The term \emph{part} introduced above usually refers to {}``word'',
but it may also refer to {}``digit'' {[}of a number{]}, {}``component''
{[}of a tuple or vector{]}, as well as group thereof. If, for example
\emph{x}, \emph{y}, \emph{z}, \emph{t} are 1-digit, 2-digit, 3-digit
and 4-digit values respectively, then $(x,y,z,t)=9876543210$ means
$x=0$, $y=21$, $z=543$ and $t=9876$.

Note that the {}``string notation'' and {}``number notation''
being used as synonyms does not mean that big-endian data ordering
is mandated. In order to avoid security irrelevant details, we do
not specify endianess. We nevertheless provide a {}``reference''
implementations in C++, where every octet string is considered as
a {[}generally multi-word{]} number with the first octet taken as
the least significant one. The implementation thus interprets octet
strings as numbers in little-endian order.

\subsection{Operations on word strings}

Let $(.)\mathrm{^{R}}$ denote the permutation that reverses the word
order of a non-empty word string. For example, for $w=8$, 

\texttt{
\[
\mathtt{0x0123ABCD}\mathrm{^{R}}=\mathtt{0xCDAB2301}
\]
}

Let $(.)\mathrm{^{S}}$ denote the permutation that swaps the high
order and low order halves of every word of a non-empty word string.
For example, for $w=8$, 

\texttt{
\[
\mathtt{0x0123ABCD}\mathrm{^{S}}=\mathtt{0x1032BADC}
\]
}

The operator $\oplus$ on word strings denote word-wise application
of $\oplus$. For example, 

\[
(a_{0},a_{1},a_{2},...)\oplus(b_{0},b_{1},b_{2},...)=(a_{0}\oplus b_{0},a_{1}\oplus b_{1},a_{2}\oplus b_{2}...)
\]

Unless otherwise said, operators $\boxplus$ and $\boxminus$ on word
strings denote word-wise modular addition and subtraction, respectively.
For example, 

\[
(a_{0},a_{1},a_{2},\ldots)\boxplus(b_{0},b_{1},b_{2},\ldots)=(a_{0}\boxplus b_{0},a_{1}\boxplus b_{1},a_{2}\boxplus b_{2},\ldots)
\]

\[
(a_{0},a_{1},a_{2},\ldots)\boxminus(b_{0},b_{1},b_{2},\ldots)=(a_{0}\boxminus b_{0},a_{1}\boxminus b_{1},a_{2}\boxminus b_{2},\ldots)
\]

Let $\bar{(.)}$ denote the word-wise application of the inversion
operator $\bar{(.)}$ on a word string. For example, 

\[
\overline{(a,b,c,...)}=(\bar{a},\bar{b},\bar{c},...)
\]

Given word strings \emph{E} and \emph{X} of the same length, let $\frac{E}{X}$
denote the word-wise application of the inversion operator $\frac{e}{x}$
on every word \emph{x} of \emph{X} with the index-matching word of
\emph{E} taken as the {[}right{]} unit element \emph{e}. For example,

\[
\frac{(e_{1},e_{2},e_{3},\ldots)}{(x_{1},x_{2},x_{3},\ldots)}=\left(\frac{e_{1}}{x_{1}},\frac{e_{2}}{x_{2}},\frac{e{}_{3}}{x_{3}},\ldots\right)
\]

Operations on word strings are used in this article only to express
the decryption function explicitly.

\section{Specification}

This section provides details of NSABC/\emph{w}. From now on \emph{w},
the word length, must be even.

Throughout this article, \emph{X} denotes a 4\emph{w}-bit plaintext
block, \emph{Y} a 4\emph{w}-bit ciphertext block, \emph{Z} a 5\emph{w}-bit
key, \emph{T} a 4\emph{w}-bit secret \emph{tweak}, i.e., a value that
is used to encrypt only one block under the key, \emph{U} a \emph{w}-bit
\emph{unit key}, i.e. an additional key that generates right unit
elements for the underlying quasi-groups.

Tweaking is optional. It may be disabled by keeping \emph{T} constant
(like \emph{Z} and \emph{U}) while encrypting many blocks. When tweaking
is disabled, NSABC becomes a conventional, non-tweakable, block cipher.

Mathematically, the cipher is given by two functions,
\begin{lyxlist}{00.00.0000}
\item [{$\textrm{ENCRYPT}(X,Z,T,U)$,}] which encrypts \emph{X} under control
of \emph{Z}, \emph{T} and \emph{U},
\item [{$\textrm{DECRYPT}(Y,Z,T,U)$,}] which decrypts \emph{Y} under control
of \emph{Z}, \emph{T} and \emph{U},
\end{lyxlist}
satisfying the apparent relation

\[
\mathrm{DECRYPT}(\mathrm{ENCRYPT}(X,Z,T,U),Z,T,U)=X
\]

The function ENCRYPT is defined in terms of four functions:
\begin{lyxlist}{00.00.0000}
\item [{CRYPT,}] a \emph{text encryption} function that encrypts a plaintext
block using a \emph{key schedule}, a \emph{unit schedule} and a \emph{tweak
schedule}; 
\item [{KE,}] a \emph{key expansion} function, that generates the key schedule
from \emph{Z};
\item [{UE,}] a \emph{unit element} function, that generates the unit schedule
from \emph{U}; and
\item [{TE,}] a \emph{tweak expansion} function, that generates the tweak
schedule from \emph{T}.
\end{lyxlist}
\begin{algorithm}[h]
\caption{\label{alg:Function-ENCRYPT}Function ENCRYPT}

\begin{description}
\item [{Input}]~

\begin{lyxlist}{00.00.0000}
\item [{\emph{X}}] 4\emph{w}-bit plaintext block
\item [{\emph{Z}}] 5\emph{w}-bit key
\item [{\emph{T}}] 4\emph{w}-bit tweak
\item [{\emph{U}}] \emph{w}-bit unit key
\end{lyxlist}
\item [{Output}]~

\begin{lyxlist}{00.00.0000}
\item [{\emph{Y}}] 4\emph{w}-bit ciphertext block
\end{lyxlist}
\item [{Relation}]~

$\mathrm{ENCRYPT}(X,Z,T,U)=\mathrm{CRYPT}(X,\mathrm{\, KE}(Z),\mathrm{UE}(U),\mathrm{TE}(T))$\end{description}
\end{algorithm}

An explicit relation for ENCRYPT is given in Algorithm~\ref{alg:Function-ENCRYPT}.
An explicit relation for DECRYPT is given in Algorithm~\ref{alg:Function-DECRYPT}.

Mechanically, encryption is performed on a conceptual processor with
a 4-word \emph{text register} $(x_{0},x_{1},x_{2},x_{3})$, a 5-word
\emph{key register} $(z_{0},z_{1},z_{2},z_{3},z_{4})$, a 4-word \emph{tweak
register} $(t_{0},t_{1},t_{2},t_{3})$ and a word \emph{unit register}
\emph{u}. The key register is initially loaded with the key \emph{Z}.
The tweak register is initially loaded with the tweak \emph{T}. The
unit register is initially loaded with the unit key \emph{U}. The
text register is initially loaded with the plaintext block \emph{X}
and finally it contains the ciphertext block \emph{Y}. 

The concrete, vector, notation here specifies the order of words so,
for example, $x_{0}$ is initially loaded with the least significant
word of \emph{X} and finally it contains the least significant word
of \emph{Y}.

\subsection{Text encryption }

The text register $(x_{0},x_{1},x_{2},x_{3})$ is initially loaded
with the plaintext block \emph{X} and finally it contains the ciphertext
block \emph{Y}.

Text encryption proceeds in 32 rounds of operations. A round is of
either \emph{type A} or \emph{type B}. The rounds are arranged in
four \emph{passes}: firstly eight rounds of type A, then eight rounds
of type B, then eight rounds of type A again, finally eight rounds
of type B again.

\begin{figure}
\includegraphics{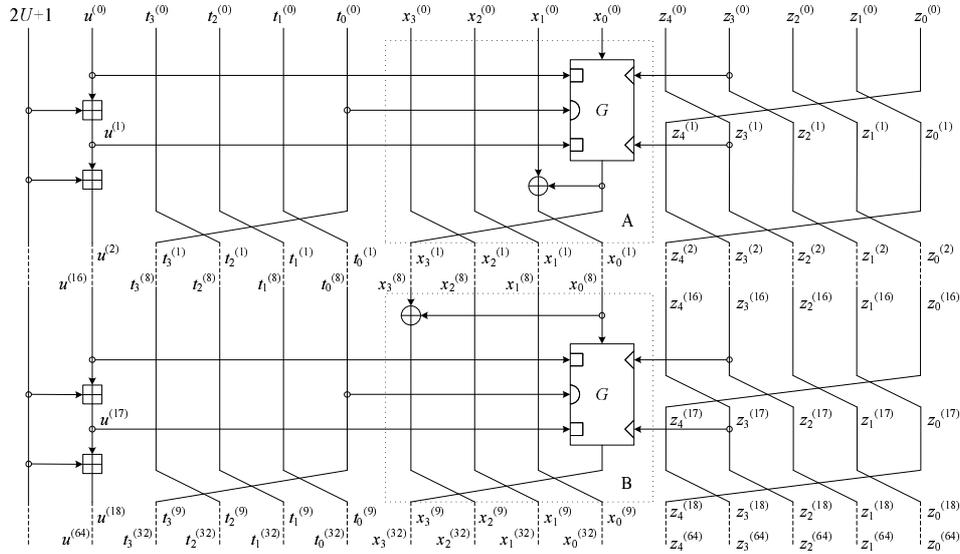}

\caption{\label{fig:Rounds} Representative rounds.}
\end{figure}

For \emph{k}-th round, $0\leq k\leq31$, the text word $x_{0}$ is
permuted, i.e. it is updated by an execution unit called \emph{G-box}
that implements a permutation \emph{G} on the set of word values,
and the contents of the text word $x_{0}$ are mixed, by exclusive-or
(XOR), into an other text word that is either $x_{1}$ or $x_{3}$.
The order of operations and the target of mixing depend on the round
type:
\begin{itemize}
\item For an A-typed round (see Figure~\ref{fig:Rounds} part A), \emph{G}
applies first, then the mixing takes place and targets $x_{1}$. That
is, the contents of $x_{0}$ enters the \emph{G}-box, the output value
of the \emph{G}-box is stored back to $x_{0}$, then the contents
of $x_{0}$ and $x_{1}$ are XOR'ed and the result is stored to $x_{1}$.
The words $x_{2}$ and $x_{3}$ are left unchanged.
\item For a B-typed round (see Figure~\ref{fig:Rounds} part B), the mixing
takes place and targets $x_{3}$ first, then \emph{G} applies. That
is, the contents of $x_{0}$ and $x_{3}$ are XOR'ed and the result
is stored to $x_{3}$, then the contents of $x_{0}$ enters the \emph{G}-box,
the output value of the \emph{G}-box is stored back to $x_{0}$. The
words $x_{1}$ and $x_{2}$ are left unchanged.
\end{itemize}

Besides the text input, the \emph{G}-box also takes as its inputs
an ordered pair of \emph{w}-bit \emph{key words} $(K{}_{2k},K{}_{2k+1})$
(depicted by $\triangleleft$ in Fig.~\ref{fig:Rounds} and \ref{fig:Permutation-G}),
an ordered pair of \emph{w}-bit \emph{unit words} $(L_{2k},L_{2k+1})$
(depicted by $^{\square}$ in Fig.~\ref{fig:Rounds} and \ref{fig:Permutation-G}),
and a \emph{w}-bit \emph{tweak word} $C_{k}$ (depicted by \textsf{\scriptsize D}
in Fig.~\ref{fig:Rounds} and \ref{fig:Permutation-G}). The details
on how key words, unit words and tweak words are generated and used
will be given in the subsequent subsections.

The encryption round is completed with a rotation by one word toward
the least significant word on the text register, i.e. the text register
is modified by simultaneous loading the word $x_{0}$ with the contents
of the word $x_{1}$, $x_{1}$ with the contents of $x_{2}$, $x_{2}$
with the contents of $x_{3}$, and $x_{3}$ with the contents of $x_{0}$.

\begin{algorithm}
\caption{\label{alg:Function-CRYPT}Function CRYPT (text encryption)}

\begin{description}
\item [{Input}]~

\begin{lyxlist}{00.00.0000}
\item [{\emph{X}}] 4\emph{w}-bit plaintext block
\item [{\emph{K}}] 64\emph{w}-bit key schedule
\item [{\emph{L}}] 64\emph{w}-bit unit schedule
\item [{\emph{C}}] 32\emph{w}-bit tweak schedule
\end{lyxlist}
\item [{Output}]~

\begin{lyxlist}{00.00.0000}
\item [{\emph{Y}}] 4\emph{w}-bit ciphertext block
\end{lyxlist}
\item [{Pseudo-code}]~\end{description}
\begin{lyxcode}
\textrm{$(x_{0},x_{1},x_{2},x_{3})\leftarrow X$}

for~$k\leftarrow0,1,2,\ldots,31$~loop

~~if~$0\leqslant k<8\;\vee\;16\leqslant k<24$~then

~~~~$x_{0}\leftarrow G(x_{0},(K_{2k},K_{2k+1}),(L_{2k},L_{2k+1}),C_{k})$

~~~~$x_{1}\leftarrow x_{1}\oplus x_{0}$~

~~elsif~$8\leqslant k<16\;\vee\;24\leqslant k<32$~then

~~~~$x_{3}\leftarrow x_{3}\oplus x_{0}$

~~~~$x_{0}\leftarrow G(x_{0},(K_{2k},K_{2k+1}),(L_{2k},L_{2k+1}),C_{k})$

~~end~if

~~\textrm{$(x_{0},x_{1},x_{2},x_{3})\leftarrow(x_{1},x_{2},x_{3},x_{0})$}

end~loop

\textrm{$Y\leftarrow(x_{0},x_{1},x_{2},x_{3})$}\end{lyxcode}
\begin{description}
\item [{Relations}]~\end{description}
\begin{lyxcode}
$Y=(x_{0}^{(32)},x_{1}^{(32)},x_{2}^{(32)},x_{3}^{(32)})$

For~$0\leqslant k<8\;\vee\;16\leqslant k<24$:

~~$x_{0}^{(k+1)}=x_{1}^{(k)}\oplus g^{(k)}$

~~$x_{1}^{(k+1)}=x_{2}^{(k)}$

~~$x_{2}^{(k+1)}=x_{3}^{(k)}$

~~$x_{3}^{(k+1)}=g^{(k)}$

For~$8\leqslant k<16\;\vee\;24\leqslant k<32$:

~~$x_{0}^{(k+1)}=x_{1}^{(k)}$

~~$x_{1}^{(k+1)}=x_{2}^{(k)}$

~~$x_{2}^{(k+1)}=x_{3}^{(k)}\oplus x_{0}^{(k)}$

~~$x_{3}^{(k+1)}=g^{(k)}$~

For~$0\leqslant k<32$:

~~$g^{(k)}=G(x_{0}^{(k)},(K_{2k},K_{2k+1}),(L_{2k},L_{2k+1}),C_{k})$~

$(x_{0}^{(0)},x_{1}^{(0)},x_{2}^{(0)},x_{3}^{(0)})=X$

$(K_{0},K_{1},K_{2},\ldots,K_{63})=K$

$(L_{0},L_{1},L_{2},\ldots,L_{63})=L$

$(C_{0},C_{1},C_{2},\ldots,C_{31})=C$\end{lyxcode}
\end{algorithm}

\subsection{Tweak schedule}

The tweak register $(t_{0},t_{1},t_{2},t_{3})$ is initially loaded
with the tweak \emph{T}. The tweak words are generated in 32 rounds
of operations.

For \emph{k}-th round, $0\leq k\leq31$, the value of the word $t_{0}$
of the tweak register is taken as the tweak word $C_{k}$ {[}which
enters the \emph{G}-box in the \emph{k}-th encryption round{]}. Then,
similarly to the text register, the tweak register is rotated by one
word toward the least significant word (see Figure~\ref{fig:Rounds}).

\begin{algorithm}

\caption{\label{alg:Function-TE}Function TE (tweak expansion) }

\begin{description}
\item [{Input}]~

\begin{lyxlist}{00.00.0000}
\item [{\emph{T}}] 4\emph{w}-bit tweak
\end{lyxlist}
\item [{Output}]~

\begin{lyxlist}{00.00.0000}
\item [{\emph{C}}] 32\emph{w}-bit tweak schedule
\end{lyxlist}
\item [{Pseudo-code}]~\end{description}
\begin{lyxcode}
$(t_{0},t_{1},t_{2},t_{3})\leftarrow T$

for~$k\leftarrow0,1,2,\ldots,31$~loop

~~$C_{k}\leftarrow t_{0}$

~~$(t_{0},t_{1},t_{2},t_{3})\leftarrow(t_{1},t_{2},t_{3},t_{0})$

end~loop~\end{lyxcode}
\begin{description}
\item [{Relations}]~\end{description}
\begin{lyxcode}
$C=(C_{0},C_{1},C_{2},\ldots,C_{31})$

For~$0\leqslant k<32$:~

~~$C_{k}=t_{0}^{(k)}$

~~$t_{0}^{(k+1)}=t_{1}^{(k)}$

~~$t_{1}^{(k+1)}=t_{2}^{(k)}$

~~$t_{2}^{(k+1)}=t_{3}^{(k)}$

~~$t_{3}^{(k+1)}=t_{0}^{(k)}$

$(t_{0}^{(0)},t_{1}^{(0)},t_{2}^{(0)},t_{3}^{(0)})=T$~\end{lyxcode}
\end{algorithm}

NOTE. For $T_{3}T_{2}T_{1}T_{0}=T$, the tweak schedule is

\[
\mathrm{TE}(T)=(T_{0},T_{1},T_{2},T_{3},T_{0},T_{1},T_{2},T_{3},\ldots,T_{0},T_{1},T_{2},T_{3})
\]

\subsection{Key schedule}

The key register $(z_{0},z_{1},z_{2},z_{3},z_{4})$ is initially loaded
with the key \emph{Z}. The key words are generated in 64 rounds of
operations. 

For \emph{k}-th round, $0\leq k\leq63$, the value of the word $z_{3}$
of the key register is taken as the key word $K_{k}$ {[}which enters
the \emph{G}-box of the \emph{k}/2-th encryption round as the first
key word if \emph{k} is even, or as the second key word if \emph{k}
is odd{]}. The key register is then rotated by one word toward the
least significant word. The rotation is similar to that on the text
register and the tweak register. (See Figure~\ref{fig:Rounds}.)

\begin{algorithm}

\caption{\label{alg:Function-KE}Function KE (key expansion)}

\begin{description}
\item [{Input}]~

\begin{lyxlist}{00.00.0000}
\item [{\emph{Z}}] 5\emph{w}-bit key
\end{lyxlist}
\item [{Output}]~

\begin{lyxlist}{00.00.0000}
\item [{\emph{K}}] 64\emph{w}-bit key schedule
\end{lyxlist}
\item [{Pseudo-code}]~\end{description}
\begin{lyxcode}
$(z_{0},z_{1},z_{2},z_{3},z_{4})\leftarrow Z$

for~$k\leftarrow0,1,2,\ldots,63$~loop

~~$K_{k}\leftarrow z_{3}$

~~$(z_{0},z_{1},z_{2},z_{3},z_{4})\leftarrow(z_{1},z_{2},z_{3},z_{4},z_{0})$

end~loop\end{lyxcode}
\begin{description}
\item [{Relations}]~\end{description}
\begin{lyxcode}
$K=(K_{0},K_{1},K_{2},\ldots,K_{63})$

For~$0\leqslant k<64$:~

~~$K_{k}=z_{3}^{(k)}$

~~$z_{0}^{(k+1)}=z_{1}^{(k)}$

~~$z_{1}^{(k+1)}=z_{2}^{(k)}$

~~$z_{2}^{(k+1)}=z_{3}^{(k)}$

~~$z_{3}^{(k+1)}=z_{4}^{(k)}$

~~$z_{4}^{(k+1)}=z_{0}^{(k)}$

$(z_{0}^{(0)},z_{1}^{(0)},z_{2}^{(0)},z_{3}^{(0)},z_{4}^{(0)})=Z$\end{lyxcode}
\end{algorithm}

NOTE. For $Z_{4}Z_{3}Z_{2}Z_{1}Z_{0}=Z$, the key schedule is

\[
\mathrm{KE}(Z)=(Z_{3},Z_{4},Z_{0},Z_{1},Z_{2},Z_{3},Z_{4},Z_{0},Z_{1},Z_{2},\ldots,Z_{3},Z_{4},Z_{0},Z_{1})
\]

\subsection{Unit schedule}

The unit register \emph{u} is initially loaded with the unit key \emph{U}.

Unit words are generated in 64 rounds of operations. 

For \emph{k}-th round, $0\leq k\leq63$, the value of the unit register
\emph{u} is taken as the unit word $L_{k}$ {[}which, similarly to
the key word $K_{k}$, enters the \emph{G}-box of \emph{k}/2-th encryption
round as the first unit word if \emph{k} is even, or as the second
unit word if \emph{k} is odd{]}. The register is then added modulo
$2^{w}$ by the {[}key-dependent{]} constant $2U\boxplus1$ to become
ready for the next round. (See Figure~\ref{fig:Rounds}.)

\begin{algorithm}

\caption{\label{alg:Function-UE}Function UE (unit element)}

\begin{description}
\item [{Input}]~

\begin{lyxlist}{00.00.0000}
\item [{\emph{U}}] \emph{w}-bit unit key
\end{lyxlist}
\item [{Output}]~

\begin{lyxlist}{00.00.0000}
\item [{\emph{L}}] 64\emph{w}-bit unit schedule
\end{lyxlist}
\item [{Pseudo-code}]~\end{description}
\begin{lyxcode}
$u\leftarrow U$

for~$k\leftarrow0,1,2,\ldots,63$~loop

~~$L_{k}\leftarrow u$

~~$u\leftarrow u\boxplus2U\boxplus1$

end~loop~\end{lyxcode}
\begin{description}
\item [{Relations}]~\end{description}
\begin{lyxcode}
$L=(L_{0},L_{1},L_{2},\ldots,L_{63})$

For~$k=0,1,2,\ldots,63$:

~~$L_{k}=u^{(k)}$

~~$u^{(k+1)}=u^{(k)}\boxplus2U\boxplus1$

$u^{(0)}=U$

Or,~equivalently,~for~every~$k$:

~~$u^{(k)}=U\odot k$~\end{lyxcode}
\end{algorithm}

NOTE. Given unit key \emph{U}, the unit schedule is

\[
\mathrm{UE}(U)=(U,3U\boxplus1,5U\boxplus2,7U\boxplus3,\ldots,127U\boxplus63)
\]

\subsection{G-box}

\begin{figure}
{\footnotesize \includegraphics{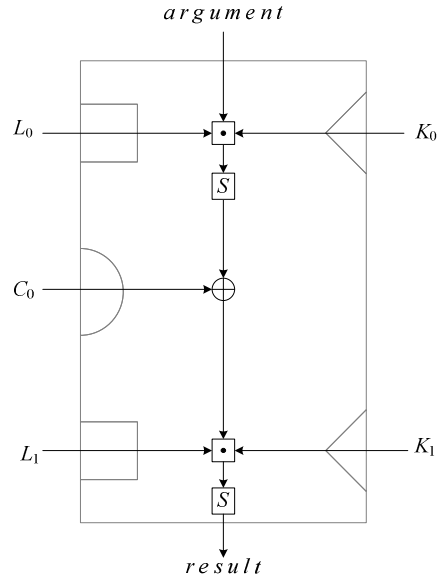}}{\footnotesize \par}

\caption{\label{fig:Permutation-G}Permutation G.}
\end{figure}

The \emph{G}-box implements a permutation \emph{G} (see Figure~\ref{fig:Permutation-G})
that takes as argument a text word and is parametrized by an ordered
pair of key words $(K_{0},K_{1})$, an ordered pair of unit words
$(L_{0},L_{1})$ and a tweak word $C_{0}$ to return a text word as
the result. The \emph{G}-box operates on a word register that initially
contains the argument and finally contains the result. The \emph{G}-box
proceeds in two rounds, each consisting of an operation $\underset{e}{\boxdot}$
followed by a half-word swap S. The two rounds are separated by an
exclusive-or (XOR) operation.

For the first round, the operation $\underset{e}{\boxdot}$ takes
the contents of the register as the left operand, $K_{0}$ as the
right operand, and $L_{0}$ as its right unit element. The result
is stored back to the register. The register is then modified by operation
S, i.e. swapping the contents of its high and low order halves.

For the inter-round XOR operation, the register is modified by XOR'ing
its contents with the tweak word $C_{0}$ and storing the result back
to it.

For the second round, the register is processed similarly to the first
round with $K_{1}$ and $L_{1}$ being used instead of $K_{0}$ and
$L_{0}$, respectively.

\begin{algorithm}
\caption{\label{alg:Permutation-G}Permutation G}

\begin{description}
\item [{Input}]~

\begin{lyxlist}{00.00.0000}
\item [{\emph{x}}] \emph{w}-bit text word
\item [{$(K_{0},K_{1})$}] pair of \emph{w}-bit key words
\item [{$(L_{0},L_{1})$}] pair of \emph{w}-bit unit words
\item [{$C_{0}$}] \emph{w}-bit tweak word
\end{lyxlist}
\item [{Output}]~

\begin{lyxlist}{00.00.0000}
\item [{\emph{y}}] \emph{w}-bit text word
\end{lyxlist}
\item [{Pseudo-code}]~\end{description}
\begin{lyxcode}
$x\leftarrow x\underset{L_{0}}{\boxdot}K_{0}$

$x\leftarrow x^{\mathrm{S}}$

$x\leftarrow x\oplus C_{0}$

$x\leftarrow x\underset{L_{1}}{\boxdot}K_{1}$

$x\leftarrow x^{\mathrm{S}}$

$y\leftarrow x$\end{lyxcode}
\begin{description}
\item [{Relation}]~\end{description}
\begin{lyxcode}
$G(x,(K_{0},K_{1}),(L_{0},L_{1}),C_{0})=(((x\underset{L_{0}}{\boxdot}K_{0})^{\mathrm{S}}\oplus C_{0})\underset{L_{1}}{\boxdot}K_{1})\mathrm{^{S}}$\end{lyxcode}
\end{algorithm}

\paragraph*{NOTES}
\begin{enumerate}
\item The cipher uses 64 distinct instances from the family of operations
$\underset{e}{\boxdot}$.
\item Alternatively, it may be seen as using 64 identical instances of the
single operation $\boxdot$ or $\odot$, but operands and result of
each instance are {}``biased'' by adding or subtracting the constant
$L_{0}$ (or $L_{1}$) that is specific to the instance, and furthermore,
being seen as $\odot$, the left operand enters and the result leaves
it in altered sign.
\item Like Skipjack, the \emph{G}-box permutes $K_{0}$ (or $K_{1}$) while
keeping $x$ and other parameters fixed. Unlike Skipjack, the \emph{G}-box
doesn't permute the word $(\mathrm{Hi}(K_{0}),\mathrm{Lo}(K_{1}))$
where $\mathrm{Hi}(.)$ and $\mathrm{Lo}(.)$ stand for the high and
the low order half respectively.
\item Unlike Skipjack, diffusion in the \emph{G}-box is incomplete, i.e.
not every input bit affects all output bits. Indeed, the \emph{v}-th
bit of the argument, with $v>w/2$, affects only all bits of the low
order half and bits \emph{v} through $w-1$ of the result; bits $w/2$
through $v-1$ remain unaffected.
\item If $(K_{0},K_{1})=(L_{0},L_{1})\:\wedge\: C_{0}=0$ then \emph{G}
becomes the identity.
\end{enumerate}

\subsection{Decryption}

Decryption can be easily derived from encryption. Namely, if

\[
\mathrm{\mathit{Y}=CRYPT}(X,\mathrm{KE}(Z),\mathrm{UE}(U),\mathrm{TE}(T))
\]

then it immediately follows that

\begin{algorithm}
\caption{\label{alg:Function-DECRYPT}Function DECRYPT}

\begin{description}
\item [{Input}]~

\begin{lyxlist}{00.00.0000}
\item [{\emph{Y}}] 4\emph{w}-bit ciphertext block
\item [{\emph{Z}}] 5\emph{w}-bit key
\item [{\emph{T}}] 4\emph{w}-bit tweak
\item [{\emph{U}}] \emph{w}-bit unit key
\end{lyxlist}
\item [{Output}]~

\begin{lyxlist}{00.00.0000}
\item [{\emph{X}}] 4\emph{w}-bit plaintext block
\end{lyxlist}
\item [{Relation}]~

\[
\mathrm{DECRYPT}(Y,Z,T,U)=\mathrm{CRYPT}(Y^{\mathrm{RS}},\frac{\mathrm{UE}(U)\mathrm{^{R}}}{\mathrm{KE}(Z^{\mathrm{R}})}\mathrm{,\mathrm{UE}(U)\mathrm{^{R}},TE}(T\mathrm{^{RS}}))\mathrm{^{RS}}
\]
\end{description}
\end{algorithm}

\[
\mathrm{CRYPT}(Y\mathrm{^{RS}},\mathrm{\frac{UE(\mathit{U})^{R}}{KE(\mathit{Z}^{R})}},\mathrm{\mathrm{UE}(U)\mathrm{^{R}},TE}(T\mathrm{^{RS}}))=X^{\mathrm{RS}}
\]

In other words, encrypting the cipher block in reverse half-word order
($Y^{\mathrm{RS}}$) using the tweak in reverse half-word order ($T^{\mathrm{RS}}$),
the unit schedule in reverse word order ($\mathrm{UE}(U)\mathrm{^{R}}$),
and the key schedule consisting of inverse words of one expanded from
the key in reverse word order ($Z^{\mathrm{R}}$), where the inversions
are of the quasi-groups defined by the operation $\underset{e}{\boxdot}$
and each quasi-group is uniquely given by its right unit that is the
index-matching word of the encryption unit schedule in reverse word
order ($\mathrm{UE}(U)\mathrm{^{R}}$), recovers the plain block in
reverse half-word order ($X^{\mathrm{RS}}$).

\begin{figure}
{\scriptsize \includegraphics[scale=1.2]{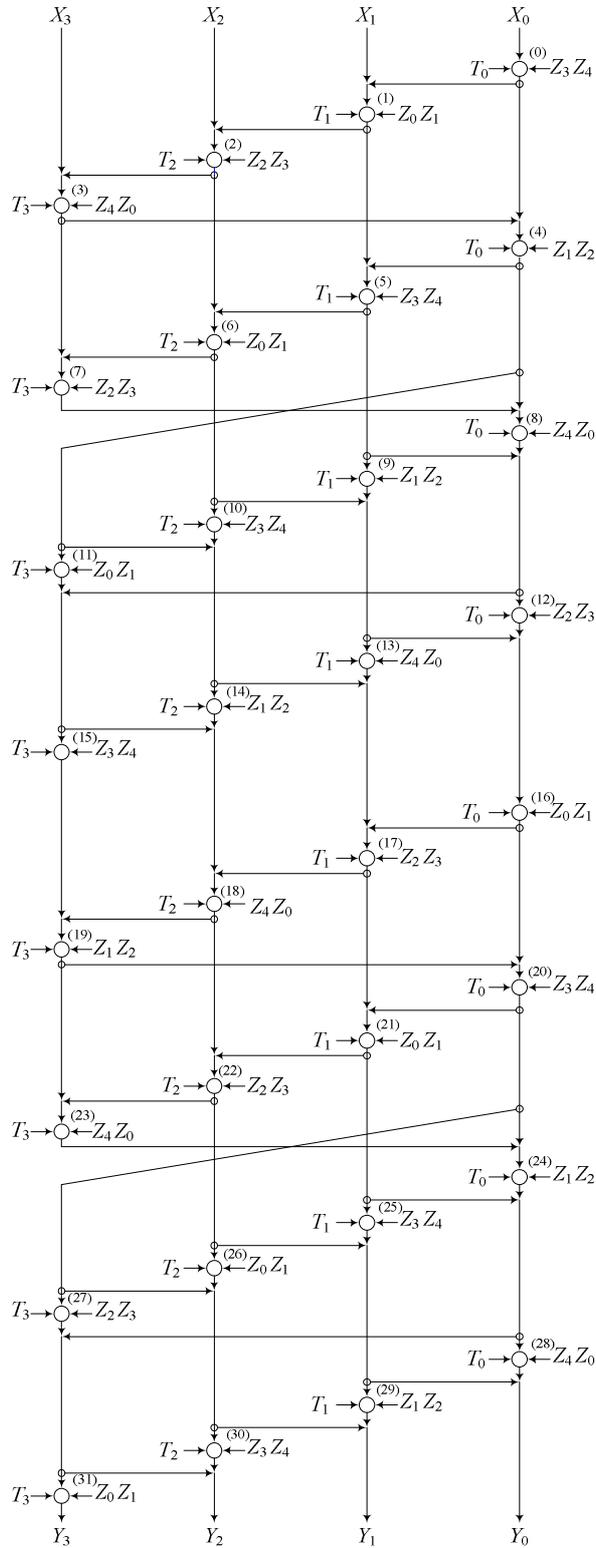}}{\scriptsize \par}

\caption{\label{fig:The-full-cipher-1}The full cipher, by {}``unrolling''.}
\end{figure}

\paragraph*{NOTES}
\begin{enumerate}
\item The full cipher is illustrated in Figure~\ref{fig:The-full-cipher-1},
where $X_{3}X_{2}X_{1}X_{0}=X$, $Y_{3}Y_{2}Y_{1}Y_{0}=Y$, $Z_{4}Z_{3}Z_{2}Z_{1}Z_{0}=Z$
and $T_{3}T_{2}T_{1}T_{0}=T$ (the unit schedule is omitted). The
figure is obtained by {}``unrolling'' (i.e. eliminating all rotations
of) the dataflow graph of the full cipher that would be obtained by
cascading the individual rounds as in Figure~\ref{fig:Rounds}.
\item The overall structure is up to word indexing identical to that of
Skipjack. The word re-indexing, which is cryptographically insignificant,
was introduced to ease description and illustration.
\item Like Skipjack, decryption is similar to encryption. To decrypt with
Skipjack, one swaps adjacent words in the cipher and the plain block
and swaps adjacent word pairs in the key. To decrypt with NSABC, one
reverses the word order, i.e., swaps the first and the last words
as well as the second first and the second last ones in the text block,
the tweak and the key. For Skipjack, one also swaps high and low order
halves of every word. For NSABC, one swaps high and low order halves
of every word but that of the key.
\item Unlike Skipjack, just swapping the words and half-words doesn't turn
encryption into decryption -- one needs to invert key words too. Thus
although ENCRYPT and DECRYPT can be expressed explicitly in terms
of CRYPT, DECRYPT cannot be expressed explicitly in terms of ENCRYPT.
\end{enumerate}

\subsection{Tweak derivation}

The 4\emph{w}-bit secret tweak \emph{T} is used to encrypt only one
block {[}under a given key \emph{Z} and unit key \emph{U}{]}. In order
to encrypt multiple blocks the tweak is derived from the block index
and a 4\emph{w}-bit {[}additional{]} key, called \emph{tweak key},
as follows. Let $T^{(j)}$ denote the tweak used to encrypt \emph{j}-th
block. For the first block ($j=0$), the tweak key is used as the
tweak directly: 

\[
T^{(0)}=\mathrm{tweak\, key}
\]

The subsequent tweak is computed from the current tweak by the recurrent
relation:

\[
T^{(j+1)}=T^{(j)}\boxplus2T^{(0)}\boxplus1
\]

or, equivalently,

\[
T^{(j)}=T^{(0)}\:\odot\: j
\]

where all operands are regarded as 4\emph{w}-bit numbers and all operators
are defined on 4\emph{w}-bit arithmetic, i.e. mod $2^{4w}$.

\paragraph*{NOTES}
\begin{enumerate}
\item The third relation, where $T^{(0)}$ conveniently designates the {[}unnamed{]}
tweak key, is meant for random access. The family of functions $\left\{ T:\, j\mapsto T^{(0)}\odot\, j\right\} $,
parametrized by the tweak key $T^{(0)}$, is not $\epsilon$-almost
2-XOR universal according to definition in \cite{LRW02}. Eventual
application of this family in the Liskov-Rivest-Wagner construction,
i.e. encryption by $\mathrm{CRYPT}(X\oplus T^{(j)},\mathrm{KE}(Z),0,\mathrm{UE}(U))\oplus T^{(j)}$,
is therefore impossible.
\item For efficient random access, applications may opt to use non-flat
spaces of the block index \emph{j}. For example, an application that
encrypts relational databases may define the index in the format $j=j_{4\,}j_{3}\, j_{2}\, j_{1}\, j_{0}$,
where $j_{4}$ is database number, $j_{3}$ is table number within
the database, $j_{2}$ is row number within the table, $j_{1}$ is
field number within the row and $j_{0}$ is block number within the
field.
\item Tweaking must be disabled when the cipher is used as a permutation,
i. e. to generate a sequence of unique numbers.
\item Tweaking should be enabled in all other modes of operation. For example,
a non-tweakable block cipher can generate a sequence of independent
numbers by encrypting a counter block in Cipher Block Chaining (CBC)
mode; NSABC can generate a similar sequence with virtually the same
cycle length by encrypting a constant block in a {}``tweaked CBC''
mode.
\end{enumerate}

\section{Example}

An encipherment in NSABC/16 with

\[
\begin{array}{c}
X=\mathtt{0x0123456789ABCDEF}\\
Z=\mathtt{0x88880777006600050000}\\
T=\mathtt{0x0001002203334444}\\
U=\mathtt{0x1998}
\end{array}
\]

results in

\[
Y=\mathtt{0x88B14E700F51921E}
\]

Table~\ref{tab:Contents-of-registers} lists states of the {[}conceptual{]}
processor during the encipherment, i.e. the contents of all registers
at the start of round \emph{k} for \emph{k} = 0, 1, 2, ..., 64 for
key schedule and unit schedule, and \emph{k} = 0, 1, 2, ..., 32 for
tweak schedule and text encryption. The start of round 64 (32) conveniently
means the end of round 63 (31), which is that of the entire algorithm.

\begin{table}
\caption{\label{tab:Contents-of-registers}Processor states during an encipherment
by NSABC/16}

\begin{lyxcode}
{\footnotesize ===~====~====================~===~================~================}{\footnotesize \par}

{\footnotesize{}~~~~Unit~~~~~~~~~Key~register~~~~~~~Tweak~register~~~~Text~register~}{\footnotesize \par}

{\footnotesize{}~~k~~~~u~~~z4~~z3~~z2~~z1~~z0~~~k~~~t3~~t2~~t1~~t0~~~x3~~x2~~x1~~x0}{\footnotesize \par}

{\footnotesize ===~====~====================~===~================~================}{\footnotesize \par}

{\footnotesize{}~~0~1998~88880777006600050000~~~0~0001002203334444~0123456789ABCDEF}{\footnotesize \par}

{\footnotesize{}~~1~4CC9~00008888077700660005}{\footnotesize \par}

{\footnotesize{}~~2~7FFA~00050000888807770066~~~1~4444000100220333~388401234567B12F}{\footnotesize \par}

{\footnotesize{}~~3~B32B~00660005000088880777}{\footnotesize \par}

{\footnotesize{}~~4~E65C~07770066000500008888~~~2~0333444400010022~1E90388401235BF7}{\footnotesize \par}

{\footnotesize{}~~5~198D~88880777006600050000}{\footnotesize \par}

{\footnotesize{}~~6~4CBE~00008888077700660005~~~3~0022033344440001~60AC1E903884618F}{\footnotesize \par}

{\footnotesize{}~~7~7FEF~00050000888807770066}{\footnotesize \par}

{\footnotesize{}~~8~B320~00660005000088880777~~~4~0001002203334444~499160AC1E907115}{\footnotesize \par}

{\footnotesize{}~~9~E651~07770066000500008888}{\footnotesize \par}

{\footnotesize{}~10~1982~88880777006600050000~~~5~4444000100220333~C2D7499160ACDC47}{\footnotesize \par}

{\footnotesize{}~11~4CB3~00008888077700660005}{\footnotesize \par}

{\footnotesize{}~12~7FE4~00050000888807770066~~~6~0333444400010022~F1EFC2D749919143}{\footnotesize \par}

{\footnotesize{}~13~B315~00660005000088880777}{\footnotesize \par}

{\footnotesize{}~14~E646~07770066000500008888~~~7~0022033344440001~03D2F1EFC2D74A43}{\footnotesize \par}

{\footnotesize{}~15~1977~88880777006600050000}{\footnotesize \par}

{\footnotesize{}~16~4CA8~00008888077700660005~~~8~0001002203334444~273D03D2F1EFE5EA}{\footnotesize \par}

{\footnotesize{}~17~7FD9~00050000888807770066}{\footnotesize \par}

{\footnotesize{}~18~B30A~00660005000088880777~~~9~4444000100220333~1615C2D703D2F1EF}{\footnotesize \par}

{\footnotesize{}~19~E63B~07770066000500008888}{\footnotesize \par}

{\footnotesize{}~20~196C~88880777006600050000~~10~0333444400010022~A9B6E7FAC2D703D2}{\footnotesize \par}

{\footnotesize{}~21~4C9D~00008888077700660005}{\footnotesize \par}

{\footnotesize{}~22~7FCE~00050000888807770066~~11~0022033344440001~18C0AA64E7FAC2D7}{\footnotesize \par}

{\footnotesize{}~23~B2FF~00660005000088880777}{\footnotesize \par}

{\footnotesize{}~24~E630~07770066000500008888~~12~0001002203334444~B049DA17AA64E7FA}{\footnotesize \par}

{\footnotesize{}~25~1961~88880777006600050000}{\footnotesize \par}

{\footnotesize{}~26~4C92~00008888077700660005~~13~4444000100220333~851857B3DA17AA64}{\footnotesize \par}

{\footnotesize{}~27~7FC3~00050000888807770066}{\footnotesize \par}

{\footnotesize{}~28~B2F4~00660005000088880777~~14~0333444400010022~71F82F7C57B3DA17}{\footnotesize \par}

{\footnotesize{}~29~E625~07770066000500008888}{\footnotesize \par}

{\footnotesize{}~30~1956~88880777006600050000~~15~0022033344440001~D5F0ABEF2F7C57B3}{\footnotesize \par}

{\footnotesize{}~31~4C87~00008888077700660005}{\footnotesize \par}

{\footnotesize{}~32~7FB8~00050000888807770066~~16~0001002203334444~E5118243ABEF2F7C}{\footnotesize \par}

{\footnotesize{}~33~B2E9~00660005000088880777}{\footnotesize \par}

{\footnotesize{}~34~E61A~07770066000500008888~~17~4444000100220333~94FAE51182433F15}{\footnotesize \par}

{\footnotesize{}~35~194B~88880777006600050000}{\footnotesize \par}

{\footnotesize{}~36~4C7C~00008888077700660005~~18~0333444400010022~7BB394FAE511F9F0}{\footnotesize \par}

{\footnotesize{}~37~7FAD~00050000888807770066}{\footnotesize \par}

{\footnotesize{}~38~B2DE~00660005000088880777~~19~0022033344440001~11747BB394FAF465}{\footnotesize \par}

{\footnotesize{}~39~E60F~07770066000500008888}{\footnotesize \par}

{\footnotesize{}~40~1940~88880777006600050000~~20~0001002203334444~D14F11747BB345B5}{\footnotesize \par}

{\footnotesize{}~41~4C71~00008888077700660005}{\footnotesize \par}

{\footnotesize{}~42~7FA2~00050000888807770066~~21~4444000100220333~0385D14F11747836}{\footnotesize \par}

{\footnotesize{}~43~B2D3~00660005000088880777}{\footnotesize \par}

{\footnotesize{}~44~E604~07770066000500008888~~22~0333444400010022~873B0385D14F964F}{\footnotesize \par}

{\footnotesize{}~45~1935~88880777006600050000}{\footnotesize \par}

{\footnotesize{}~46~4C66~00008888077700660005~~23~0022033344440001~CB9B873B03851AD4}{\footnotesize \par}

{\footnotesize{}~47~7F97~00050000888807770066}{\footnotesize \par}

{\footnotesize{}~48~B2C8~00660005000088880777~~24~0001002203334444~D6FCCB9B873BD579}{\footnotesize \par}

{\footnotesize{}~49~E5F9~07770066000500008888}{\footnotesize \par}

{\footnotesize{}~50~192A~88880777006600050000~~25~4444000100220333~D4CF0385CB9B873B}{\footnotesize \par}

{\footnotesize{}~51~4C5B~00008888077700660005}{\footnotesize \par}

{\footnotesize{}~52~7F8C~00050000888807770066~~26~0333444400010022~779F53F40385CB9B}{\footnotesize \par}

{\footnotesize{}~53~B2BD~00660005000088880777}{\footnotesize \par}

{\footnotesize{}~54~E5EE~07770066000500008888~~27~0022033344440001~8CECBC0453F40385}{\footnotesize \par}

{\footnotesize{}~55~191F~88880777006600050000}{\footnotesize \par}

{\footnotesize{}~56~4C50~00008888077700660005~~28~0001002203334444~C93A8F69BC0453F4}{\footnotesize \par}

{\footnotesize{}~57~7F81~00050000888807770066}{\footnotesize \par}

{\footnotesize{}~58~B2B2~00660005000088880777~~29~4444000100220333~2E1A9ACE8F69BC04}{\footnotesize \par}

{\footnotesize{}~59~E5E3~07770066000500008888}{\footnotesize \par}

{\footnotesize{}~60~1914~88880777006600050000~~30~0333444400010022~8038921E9ACE8F69}{\footnotesize \par}

{\footnotesize{}~61~4C45~00008888077700660005}{\footnotesize \par}

{\footnotesize{}~62~7F76~00050000888807770066~~31~0022033344440001~D4BE0F51921E9ACE}{\footnotesize \par}

{\footnotesize{}~63~B2A7~00660005000088880777}{\footnotesize \par}

{\footnotesize{}~64~E5D8~07770066000500008888~~32~0001002203334444~88B14E700F51921E}{\footnotesize \par}

\end{lyxcode}
\end{table}

\section{Notes on implementation}

This section provides methods for efficient software implementation
for two types of environment: memory-constrained, such as embedded
computers, and memory-abundant, such as servers and personal computers.

\subsection{Memory-constrained environment}

The function ENCRYPT can be implemented without using any writeable
memory on a processor with at least 16 word registers:
\begin{itemize}
\item 4 for $(x_{0},x_{1},x_{2},x_{3})$ -- text register
\item 5 for $(z_{0},z_{1},z_{2},z_{3},z_{4})$ -- key register
\item 4 for $(t_{0},t_{1},t_{2},t_{3})$ -- tweak register
\item 1 for \emph{u} -- unit register, 
\item 1 for the constant value $2U\boxplus1$, and
\item 1 for \emph{k} -- round index.
\end{itemize}
Indeed, the schedules \emph{K}, \emph{L} and \emph{C} can vanish because
every word of them, once produced, can be consumed immediately, provided
that the functions KE, TE, UE and CRYPT are programmed to run in parallel
and synchronized with each increment of \emph{k}. Source code of this
implementation is given in Appendix A.

Unlike ENCRYPT, DECRYPT needs memory for the key schedule because
on-the-fly modular multiplicative inversion is too slow to be practical.
In this environment, modes of operation that avoid DECRYPT (i.e. ones
using ENCRYPT to decrypt) are thus preferable.

\subsection{Memory-abundant environment}

The quasi-group operation $\underset{e}{\boxdot}$ can be evaluated
by only one multiplication and one addition. Indeed, 

\[
x\underset{e}{\boxdot}z=mx\boxplus n
\]

where \emph{x} is a text word, \emph{z} key word, and

\[
m=2(z-e)\boxplus1
\]

\[
n=(2e-1)\boxtimes(z-e)
\]

So, instead of using the $(z,e)$ pairs, one may pre-compute the $(m,n)$
pairs once and use them many times.

\begin{figure}
\includegraphics{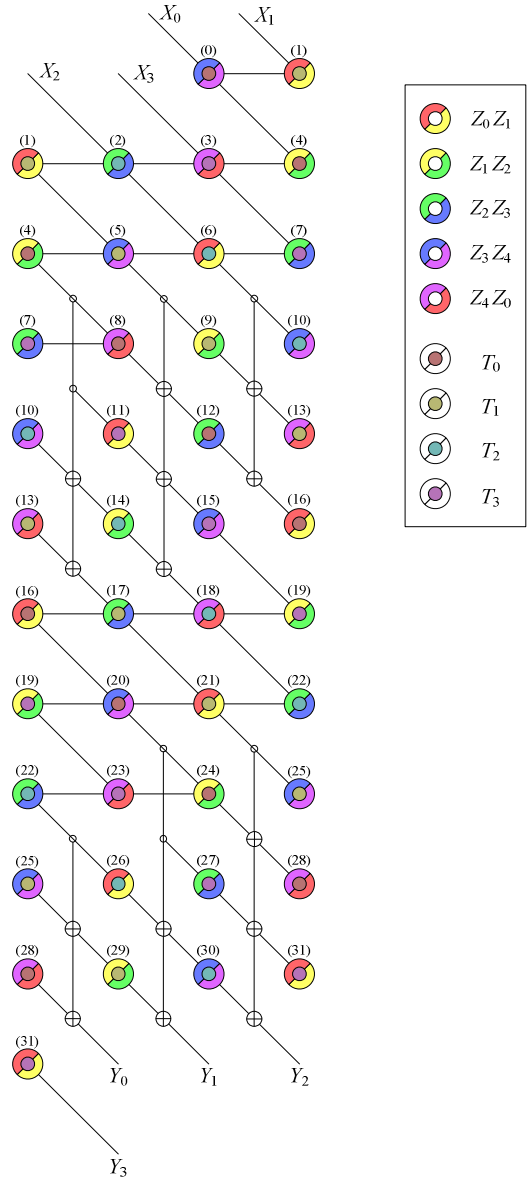}

\caption{\label{fig:The-full-cipher-2}The full cipher, by another {}``unrolling''.}

\end{figure}

The cipher is parallelizable. The following procedure executes all
32 rounds in 20 steps, of which half performing two or three parallel
evaluations of \emph{G}. Recall that $g^{(k)}$ is the result of \emph{G}
in round \emph{k}.
\begin{enumerate}
\item Compute $g^{(0)}$
\item Compute $g^{(1)}$
\item Compute $g^{(2)}$
\item Compute $g^{(3)}$
\item Compute $g^{(4)}$
\item Compute $g^{(5)}$, $g^{(11)}$ in parallel
\item Compute $g^{(6)}$, $g^{(9)}$ in parallel
\item Compute $g^{(7)}$, $g^{(10)}$, $g^{(13)}$ in parallel
\item Compute $g^{(8)}$, $g^{(14)}$ in parallel
\item Compute $g^{(12)}$, $g^{(15)}$ in parallel
\item Compute $g^{(16)}$
\item Compute $g^{(17)}$
\item Compute $g^{(18)}$
\item Compute $g^{(19)}$
\item Compute $g^{(20)}$
\item Compute $g^{(21)}$, $g^{(27)}$ in parallel
\item Compute $g^{(22)}$, $g^{(25)}$ in parallel
\item Compute $g^{(23)}$, $g^{(26)}$, $g^{(29)}$ in parallel
\item Compute $g^{(24)}$, $g^{(30)}$ in parallel
\item Compute $g^{(28)}$, $g^{(31)}$ in parallel
\end{enumerate}
The procedure becomes evident by examining the dataflow graph of the
cipher, shown in Figure~\ref{fig:The-full-cipher-2}, which is obtained
by {}``unrolling'' the one in Figure~\ref{fig:The-full-cipher-1}.
Here {}``unrolling'' means introducing a rotation so that the G-boxes
with congruent round indices (mod 3) lay on a straight line.

On a x86-64 processor in 32-bit mode ($w=32$), the procedure takes
about 256 clock cycles, i.e. 256/16 = 16 clock cycles per byte encrypted.
(The source code of this implementation is given in Appendix B.) In
64-bit mode ($w=64$), it takes about 384 clock cycles, i.e. 384/32
= 12 clock cycles per byte.

The procedure may be also coded twice, i.e. it may be run in two instances
in parallel on a single core of the processor, with the second instance
delayed by a few steps after the first, to encrypt two blocks possibly
under different tweaks and/or keys. This method has shown to be effective
for x86-64 processors in 64-bit mode, resulting in about 9 clock cycles
per byte.

\section{Conclusion}

We defined NSABC, a block cipher utilizing a group operation that
is essentially modular multiplication of machine words, a powerful
operation available on many processors.

NSABC was meant to be elegant. It uses no S-boxes or {}``magic''
constants. It uses only machine word-oriented algebraic operations.
It makes use of the simple and regular structure of Skipjack which
has become publicly known for over a decade \textemdash{} sufficient
time to be truly understood. It is elegant to be easily memorizable,
realizable and analyzable.

NSABC bases on some valuable design of a well-reputed agency in the
branch. We therefore believe that it is worth analysis and it can
withstand rigorous analysis. If this happens to be true, then we may
have a practical cipher with 256-bit blocks, allowing to encrypt enormous
amount of data under the same key, and with 320-bit keys, allowing
to protect data over every imaginable time.

In cipher design there is always a trade-off between security and
efficiency, and designers always have to ask: {}``What do we want,
a very strong and fairly fast cipher, or fairly strong but very fast?'' 

NSABC reflects the authors' view on the dilemma. If Skipjack is regarded
as very strong and just fairly fast, then NSABC may be regarded as
a design emphasizing the second aspect \textemdash{} make it very
fast, abeit just fairly strong. For \emph{w}-bit word length, NSABC
key length is 5\emph{w} bits, optionally plus 5\emph{w} bits more,
whilst the true level of security is yet to be determined. On the
other hand, on a modern 64-bit processor it takes only 9 clock cycles
to encrypt a byte.

NSABC is thus fast to be comparable to every modern block cipher.

\specialsection*{APPENDIX A. A reference implementation of NSABC/32 \textemdash{}
ENCRYPT only}

\begin{lstlisting}[basicstyle={\scriptsize\ttfamily},numbers=left,numberstyle={\footnotesize},tabsize=4]
typedef uint32_t word; 
//----------------------------------------------------------------------
static word o(word x, word y, word e)
{
    return 2*x*y + (1 - 2*e)*(x - y + e);
}
//----------------------------------------------------------------------
static word G(word x, word z0, word z1, word u0, word u1, word t)
{   
    x = o(x,z0,u0);
    x = _rotl(x,16);
    x ^= t;
    x = o(x,z1,u1);
    x = _rotl(x,16);
    return x;
}
//----------------------------------------------------------------------
void encrypt( word Y[4], word const X[4], word const Z[5], 
              word const T[4], word U )
{
    word
        x0 = X[0], x1 = X[1], x2 = X[2], x3 = X[3],
        z0 = Z[0], z1 = Z[1], z2 = Z[2], z3 = Z[3], z4 = Z[4],
        t0 = T[0], t1 = T[1], t2 = T[2], t3 = T[3], 
        u0 = U,
        u1 = u0 + 2*U + 1;
    for(int k = 0; k < 32; k++)
    {
        if(k & 8)         // B-round
        {
            x3 ^= x0;
            x0 = G(x0, z3, z4, u0, u1, t0);
        }
        else              // A-round
        {
            x0 = G(x0, z3, z4, u0, u1, t0);
            x1 ^= x0;
        }
        word x = x0; x0 = x1; x1 = x2; x2 = x3; x3 = x;
        word z = z0; z0 = z2; z2 = z4; z4 = z1; z1 = z3; z3 = z;
        word t = t0; t0 = t1; t1 = t2; t2 = t3; t3 = t;
        u0 = u1 + 2*U + 1;
        u1 = u0 + 2*U + 1;
    }
    Y[0] = x0; Y[1] = x1; Y[2] = x2; Y[3] = x3;
}
\end{lstlisting}

\specialsection*{APPENDIX B. An optimized implementation of NSABC/32}

{\footnotesize }
\begin{lstlisting}[basicstyle={\scriptsize\ttfamily},numbers=left,tabsize=4]
void expandkey   (word M[64], word N[64], word const Z[5], word U)
{
    word z0=Z[0], z1=Z[1], z2=Z[2], z3=Z[3], z4=Z[4];
    word u = U;
    for( int k=0; k<64; k++ )
    {
        M[k] = 2*(z3 - u) + 1;
        N[k] = (2*u - 1)*(z3 - u);
        u += 2*U + 1;
        word z=z0; z0=z1; z1=z2; z2=z3; z3=z4; z4=z;
    }
}
//----------------------------------------------------------------------
static inline 
word G( word x, word t, word m0, word m1, word n0, word n1 )
{
    x *= m0;
    x += n0;
    x = _rotl(x,16);
    x ^= t;
    x *= m1;
    x += n1;
    x = _rotl(x,16);
    return x;
}
//----------------------------------------------------------------------
void crypt( word Y[4], word const X[4], word const T[4], 
            word const M[64], word const N[64])
{   
    // Step 1
    word const g0 = G(X[0],              T[0],  M[0], M[1], N[0], N[1]);
    // Step 2
    word const g1 = G(X[1]^g0,           T[1],  M[2], M[3], N[2], N[3]);
    // Step 3
    word const g2 = G(X[2]^g1,           T[2],  M[4], M[5], N[4], N[5]);
    // Step 4
    word const g3 = G(X[3]^g2,           T[3],  M[6], M[7], N[6], N[7]);
    // Step 5
    word const g4 = G(g0^g3,             T[0],  M[8], M[9], N[8], N[9]);
    // Step 6
    word const g5 = G(g1^g4,             T[1], M[10],M[11],N[10],N[11]);
    word const g11= G(g4,                T[3], M[22],M[23],N[22],N[23]);
    // Step 7
    word const g6 = G(g2^g5,             T[2], M[12],M[13],N[12],N[13]);
    word const g9 = G(g5,                T[1], M[18],M[19],N[18],N[19]);
    // Step 8
    word const g7 = G(g3^g6,             T[3], M[14],M[15],N[14],N[15]);
    word const g10= G(g6,                T[2], M[20],M[21],N[20],N[21]);
    word const g13= G(g6^g9,             T[1], M[26],M[27],N[26],N[27]);
    // Step 9
    word const g8 = G(g4^g7,             T[0], M[16],M[17],N[16],N[17]);
    word const g14= G(g4^g10,            T[2], M[28],M[29],N[28],N[29]);
    // Step 10
    word const g12= G(g5^g8,             T[0], M[24],M[25],N[24],N[25]);
    word const g15= G(g5^g8^g11,         T[3], M[30],M[31],N[30],N[31]);
    // Step 11
    word const g16= G(g6^g9^g12,         T[0], M[32],M[33],N[32],N[33]);
    // Step 12
    word const g17= G(g4^g10^g13^g16,    T[1], M[34],M[35],N[34],N[35]);
    // Step 13
    word const g18= G(g5^g8^g11^g14^g17, T[2], M[36],M[37],N[36],N[37]);
    // Step 14
    word const g19= G(g15^g18,           T[3], M[38],M[39],N[38],N[39]);
    // Step 15
    word const g20= G(g16^g19,           T[0], M[40],M[41],N[40],N[41]);
    // Step 16
    word const g21= G(g17^g20,           T[1], M[42],M[43],N[42],N[43]);
    word const g27= G(g20,               T[3], M[54],M[55],N[54],N[55]);
    // Step 17
    word const g22= G(g18^g21,           T[2], M[44],M[45],N[44],N[45]);
    word const g25= G(g21,               T[1], M[50],M[51],N[50],N[51]);
    // Step 18
    word const g23= G(g19^g22,           T[3], M[46],M[47],N[46],N[47]);
    word const g26= G(g22,               T[2], M[52],M[53],N[52],N[53]);
    word const g29= G(g22^g25,           T[1], M[58],M[59],N[58],N[59]);
    // Step 19
    word const g24= G(g20^g23,           T[0], M[48],M[49],N[48],N[49]);
    word const g30= G(g20^g26,           T[2], M[60],M[61],N[60],N[61]);
    Y[1]   = g20^g26^g29;
    // Step 20
    word const g28= G(g21^g24,           T[0], M[56],M[57],N[56],N[57]);
    word const g31= G(g21^g24^g27,       T[3], M[62],M[63],N[62],N[63]);
    Y[2]   = g21^g24^g27^g30;
    // Step 21
    Y[0]   = g22^g25^g28;
    Y[3]   = g31;
}
//----------------------------------------------------------------------
// Multiplicative inverse of x (mod 2**32), x odd.
// Source code by Thomas Pornin, Usenet 2009.
word inverse(word x)
{
    word y = 2 - x; // xy == 1 mod 4
    y *= 2 - x*y;   // xy == 1 mod 16
    y *= 2 - x*y;   // xy == 1 mod 256
    y *= 2 - x*y;   // xy == 1 mod 65536
    y *= 2 - x*y;   // xy == 1 mod 4294967296
    return y;
}
//----------------------------------------------------------------------
void invertkey( word      iM[64], word      iN[64], 
                word const M[64], word const N[64] )
{
    // M, N, iM, iN must not overlap!
    for(int k=0; k < 64; k++)
    {
        iM[k] = inverse( M[63-k] );
        iN[k] = - N[63-k] * iM[k];
    }
}
//----------------------------------------------------------------------
void icrypt( word X[4], word const Y[4], word const T[4], 
             word const iM[64], word const iN[64] )
{   
    word Xrs[4], Trs[4];
    Trs[0] = _rotl(T[3],16);
    Trs[1] = _rotl(T[2],16);
    Trs[2] = _rotl(T[1],16);
    Trs[3] = _rotl(T[0],16);
    Xrs[0] = _rotl(Y[3],16);
    Xrs[1] = _rotl(Y[2],16);
    Xrs[2] = _rotl(Y[1],16);
    Xrs[3] = _rotl(Y[0],16);
    crypt( Xrs, Xrs, Trs, iM, iN );
    X[0] = _rotl(Xrs[3],16);
    X[1] = _rotl(Xrs[2],16);
    X[2] = _rotl(Xrs[1],16);
    X[3] = _rotl(Xrs[0],16);
}
//----------------------------------------------------------------------
// Testing against the reference implementation
void test()
{
    int const nTimes = 10000;
    int const nRep   = 100;
    word X[4], Y[4], T[4], Z[5], M[64], N[64], iM[64], iN[64];
    for(int i=0; i<5; i++)
        Z[i] = random_word();
    for(int i=0; i<4; i++)
        T[i] = random_word();
    word U = random_word();
    // correctness of the optimized implementation
    expandkey( M, N, Z, U );
    for(int n=nTimes; n; n--)
    {
        for(int i=0; i<4; i++)
            X[i] = random_word();
        memcpy( Y, X, sizeof(X) );
        for(int m=nRep; m; m--)
        {
            encrypt( X,  X,  Z, T, U );
            crypt( Y, Y, T, M, N );
        }
        if( memcmp(Y,X,sizeof(X)) !=0 )
            cout << "crypt: incorrect encryption!" << endl;
    }
    // invertibility of the optimized implementation
    invertkey( iM, iN, M, N );
    for(int n=nTimes; n; n--)
    {
        for(int i=0; i<4; i++)
            X[i] = random_word();
        memcpy( Y, X, sizeof(X) );
        for(int m=nRep; m; m--)
            crypt( Y, Y, T, M, N );
        for(int m=nRep; m; m--)
            icrypt( Y, Y, T, iM,iN );
        if( memcmp(Y, X, sizeof(X)) !=0 )
            cout << "icrypt: incorrect decryption!" << endl;
    }
}
////////////////////////////////////////////////////////////////////////
\end{lstlisting}

\end{document}